\documentclass{emulateapj}

\begin{document}

\title{Keck NIRSPEC Radial Velocity Observations of Late-M dwarfs}

\author{Angelle Tanner\altaffilmark{1,2}, 
Russel White\altaffilmark{1},
John Bailey\altaffilmark{3},
Cullen Blake\altaffilmark{4},
Geoffery Blake\altaffilmark{5},
Kelle Cruz\altaffilmark{6},
Adam J. Burgasser\altaffilmark{7},
Adam Kraus\altaffilmark{8}}

\altaffiltext{1}{Georgia State University, Department of Astronomy, One Park Place,  Atlanta, GA, 30303, USA }
\altaffiltext{2}{Current address: Mississippi State University, Department of Physics \& Astronomy, Hilbun Hall, Starkville, MS, 39762, USA }
\altaffiltext{3}{Department of Astronomy, University of Michigan, 830 Dennison Bldg. 500 Church St. Ann Arbor, MI 48109-1042, USA}
\altaffiltext{4}{Department of Astrophysical Sciences, Princeton University, Peyton Hall, Ivy Lane, Princeton, NJ 08544, USA}
\altaffiltext{5}{Division of Geological \& Planetary Sciences, California Institute of Technology, Pasadena, CA 91125, USA}
\altaffiltext{6}{Dept. of Physics and Astronomy, Hunter College, 695 Park Avenue, NY, NY 10065 USA and
Dept. of Astrophysics, American Museum of Natural History}
\altaffiltext{7}{Center for Astrophysics and Space Science, University of California San Diego, La Jolla, CA 92093, USA}
\altaffiltext{8}{Institute for Astronomy, University of Hawaii, 2680 Woodlawn Dr., Honolulu, HI 96822, USA, Hubble Fellow}

\begin{abstract}
We present the results of an infrared spectroscopic survey of 23 late-M dwarfs with the NIRSPEC echelle spectrometer on the Keck II telescope.  Using telluric lines for wavelength calibration, we are able to achieve measurement precisions of down to 45 m/s for our late-M dwarfs over a one to four year-long baseline. Our sample contains two stars with RV variations of $>$1000 m/s. While we require more measurements to determine whether these RV variations are due to unseen planetary or stellar companions or are the result of starspots known to plague the surface of M dwarfs, we can place upper limits of $<$40 M$_J$sini on the masses of any companions around those two M dwarfs with RV variations of $<$160 m/s at orbital periods of 10-100 days. We have also measured the rotational velocities for all the stars in our late-M dwarf sample and offer our multi-order, high-resolution spectra over 2.0 to 2.4 $\micron$ to the atmospheric modeling community to better understand the atmospheres of late-M dwarfs.  
\end{abstract}

\section{Introduction}

With at least 70\% of all stars within 10 pc identified as M dwarfs (Henry et al. 2006), these low-mass stars are becoming high priority targets for extrasolar planet surveys. With measurement precisions consistently at the 1 m/s level, optical radial velocity (RV) surveys are capable of detecting super-Earth mass planets around some of our nearest M dwarf neighbors. A classic example of the success of optical RV surveys around M dwarfs being the GJ 581 system, which harbors up to six planets including a few with masses less than 10 M$_\earth${\it sini} (Vogt et al. 2010; Mayor et al. 2008). The lightest known star with a planet is the M4V dwarf, GJ 1214, which has a 6.6 M$_\earth$ planet. This companion was discovered by the Mearth transit program (Nutzman \& Charbonneau 2008) and confirmed through optical RV measurements with a velocity amplitude of 12 m/s (Charbonneau et al. 2009).
Despite M dwarfs comprising the majority of the super-Earth mass planets, they have been neglected in many of the original large scale RV programs because they are so optically faint. 
Out of the 185 M dwarfs within 10 pc, 35\% of them are fainter than V=12 and 16\% of them are fainter than V=14 (T. Henry private comm.) 
A few surveys of early- to mid-M dwarfs with optical echelle spectrometers have achieved RV precisions of down to $\sim$3 m/s and placed limits on the population of giant planets around these types of stars of 2$^{+0.03}_{-0.01}$\% for periods of 10 to 100 days (Bonfils et al. 2012;  Johnson et al. 2010; Endl et al. 2006;  Zechmeister et al. 2006). These surveys, however, did not include any M dwarfs later than M6. One solution to the faintness issue inherent to optical observations of late-M dwarfs is to move the observations to the near-infrared part of the spectrum where M dwarfs are considerably brighter than solar-type stars. In addition, the contrast between the photosphere of the star and starspots is reduced at near-infrared wavelengths. These co-rotating features can be a prominent source of measurement noise when looking for the periodic variability produced by an orbiting planet. This is especially true for both M dwarfs and young stars which can have both flares and large starspots (Setiawan et al. 2008; Huelamo et al. 2008; Prato et al. 2008; Bailey et al. 2011). 

There are now a variety of infrared echelle spectrometers commissioned on 3-10 meter telescopes including NIRSPEC on Keck II (McLean et al. 1998),
Phoenix on Gemini South (Hinkle et al. 2003), CRIRES on the VLT (Kaufl et al. 2004) and CSHELL on the IRTF (Greene et al. 1993). All of these spectrometers are capable
of producing spectral resolutions of R$>$20000 within the 1.0 to 2.5 micron wavelength region. In the near-infrared, the Earth's atmospheric spectrum is riddled with absorption features from water, carbon dioxide and methane. This allows for a natural spectral reference frame from which to measure the change in the position of the stellar absorption lines (Griffin \& Griffin 1973). Using telluric features, the CRIRES spectrometer achieved RV precisions of 5-20 m/s for bright M giant stars (Figueira et al. 2010; Seifahrt \& Kaufl 2008). Other radial velocity precisions achieved with infrared echelle spectrometers using telluric calibration include 50-300 m/s for a sample of nearby, slowly rotating early-M dwarfs (Bailey et al. 2012; Rodler et al. 2011; Crockett et. al. 2011) and 200 m/s for a sample of brown dwarfs (Blake et al. 2010; Bailey et al. 2012). A precision of 50 m/s is sufficient to detect $>$2 $M_J$ mass planets with periods of 10 days around mid-M dwarfs. These infrared RV precisions can be compared to the $\sim$ 1 m/s precision consistently demonstrated around mid-M type stars at optical wavelengths (Bonfils et al. 2011).

Finding planets around M dwarfs is important because the two competing planet formation paradigms, core accretion and gravitational instability, predict very different planet frequencies for the smallest stars: core accretion models predict relatively few planets (1\%; Kennedy \& Kenyon 2008) while gravitational instability models (e.g. Boss et al. 2006) predict frequencies similar to those of solar-type stars, perhaps 10\% or more. Because M dwarfs are, by far, the most abundant spectral type in the stellar neighborhood, a thorough census of the planetary population around these stars is advantageous for potential follow-up studies. 

Here, we present the results of an infrared spectroscopic survey of a sample of late-M dwarfs with the NIRSPEC spectrometer on
Keck II. In Section 2, we review the observing program used to collect the data as well as the data reduction and method used to measure the radial velocities of the spectra; in Section 3, we comment on the properties of the RV measurements for those stars with multiple measurements and stellar rotational velocities for the entire sample. 
In Section 4, we discuss the implications of these measurements and the future of planet searches with the NIRSPEC instrument. 

\section{Observations and Analysis}

We have observed a sample of 23 late-M dwarfs over a spectral type range of M6.0 to M9.0 (see Table~\ref{msample}) with the NIRSPEC spectrometer on the W. H. Keck II telescope (McLean et al. 1998). To avoid close binaries that would impede RV planet detections, these stars
were chosen from high spatial resolution, adaptive optics surveys of nearby late-M dwarfs (Close et al. 2003; Seigler et al. 2003). 
Using the photometric distance relationship from Henry et al. (2006), we estimate the range in distances for this sample of M dwarfs to be 4 to 50 parsecs based on their V-K colors and spectral types. Four of the M dwarfs have photometric distances placing them within 10 parsecs of the Sun. Seven of the M dwarfs were observed three or more times over a four year time span from 2006 June  to 2010 July.  GJ 752 A and B were added to our sample in mid-2009 due to the announcement of an astrometric planet discovered around GJ 752 B (Pravdo \& Shaklan 2009). NIRSPEC data for these two stars was collected by multiple observers during unrelated NIRSPEC runs with the same instrumental setup. Additional K and early-M dwarfs (GJ 628, GJ 725 A, GJ 725 B and GJ 752 A, see Table~\ref{sample}) were observed as RV standards and to determine the infrared RV precision as a function of spectral type. Finally, a set of five A type stars were observed for determining the wavelengths solution of the instrument.

We collected near-infrared spectra using the NIRSPEC-7 (2.22 $\mu$m) passband in echelle mode with the 3 pixel slit (0.432$\arcsec$) in combination with an echelle angle of 62.65 and grating angle of 35.50 degrees. This configuration results in a wavelength coverage of 1.99 to 2.39 $\micron$ and a spectral resolution of R$\sim$24000 (see Figure~\ref{allorders}). 
Each epoch of observation included nodding the star along the slit to allow for subtraction of two images to remove background emission prior to spectral extraction. The integration times for the targets ranged from a few minutes per nod for the A stars and 30 minutes per nod for the late-M science targets. From the echelle data, which covers seven full spectral orders, we determine our radial velocities from the 33rd order on the focal plane covering the spectral range of 2.28-2.32 $\micron$. This wavelength region is ideal for RV measurements because it is permeated by telluric absorption features as well as the CO bandhead prominent in M dwarf spectra. 

\subsection{Spectral Extraction and Modeling}

The data reduction and spectral analysis are described in detail in Bailey et al. (2012). Here we highlight the primary steps in the data reduction as well as any differences between
our analysis and that presented in Bailey et al. (2012). After flat fielding and nod-pair sky sub-traction, the 1-D spectra are extracted from the NIRSPEC images using the 'optimal extraction' method first proposed by Horne (1986). This method constructs spectra by summing weighted pixel values across the spectral profile, where the weights are based on noise statistics. Although the technique works well at removing deviant pixel values caused by such things as bad pixels and cosmic ray events for high signal-to-noise spectra, occasional features sometimes remained in the lower signal-to-noise spectra of the faintest stars observed. These features appear as large positive or negative values of 1-2 pixels width and likely result from a poorly determined spectral profile at that location, perhaps caused by a larger than normal number deviant pixel values; this appears to be common for older generation infrared arrays. In these cases, these features were identified by eye in the spectra and the values were replaced by the median value of neighboring values. For each pair of observations, the spectra at each of the two nod positions are analyzed separately, providing two RV measurements per epoch.
 
Each observed 1-D NIRSPEC spectrum is modeled as the composite of the intrinsic, rotationally broadened spectrum of the star and the telluric spectrum convolved with the instrumental profile. The telluric spectrum is extrapolated from spectral observations of the Sun (Livingston \& Wallace 1991) while the intrinsic spectrum of each of the late-M dwarfs is extrapolated from the NextGen models, which account for the star's effective temperature and surface gravity (Hauschildt et al. 1999). An initial estimate of the wavelength
calibration and spectral response function are determined from a set of A star calibrators observed during the same night as the science targets. A stars are ideal spectral calibrators because the spectrum of the star is featureless in the wavelength region used to estimate the radial velocities. 
To determine the initial values of the wavelength solution and instrumental spectral profile, 
the model telluric spectrum is fitted to the observed spectrum of the A star by minimizing the $\chi^2$ value with the amoeba downhill simplex algorithm (Press et al. 1992). The fitting process involves four free parameters: three for the quadratic wavelength solution and one for the gaussian FWHM of the instrumental spectral profile. 

With initial guesses taken from the known properties of the spectrograph determined by previous NIRSPEC spectrograph programs, all four free parameters are allowed to vary until the best fit is achieved. The four parameters characterizing the instrument are then used as initial guesses for finding the best fitting model spectrum of the science targets. There are ten free parameters involved when modeling the observed M dwarf spectra: three for the quadratic wavelength solution, and one each for the instrumental spectral profile, limb darkening, RV, {\it vsini}, and airmass. There are also two scaling factors for the relative depth of the telluric and synthetic M dwarf spectrum. When available we use published values of the {\it vsini} for the initial guess during the fitting process (see Table~\ref{rms}). 

\begin{figure}[ht]
\epsscale{1.0}
\plottwo{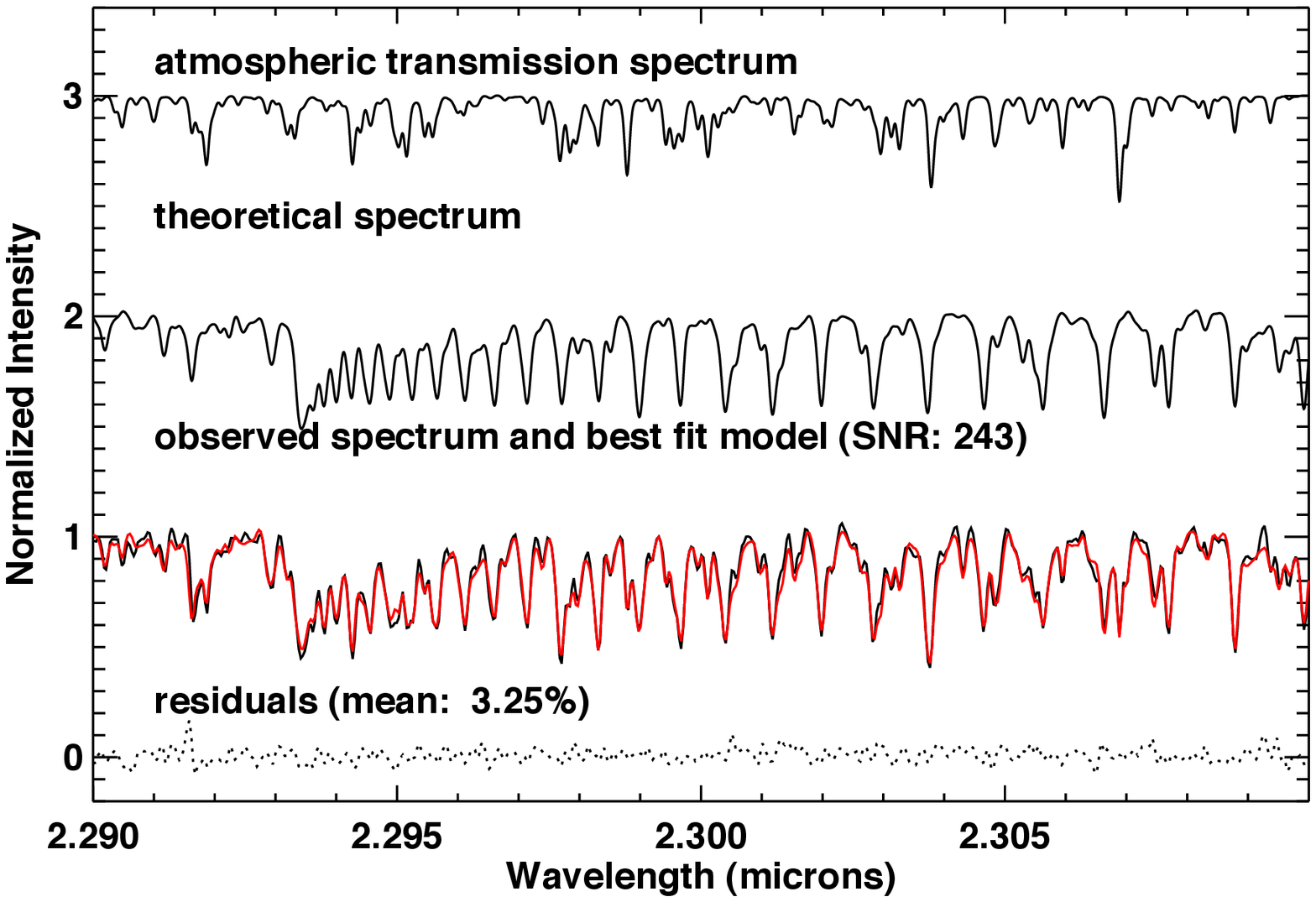}{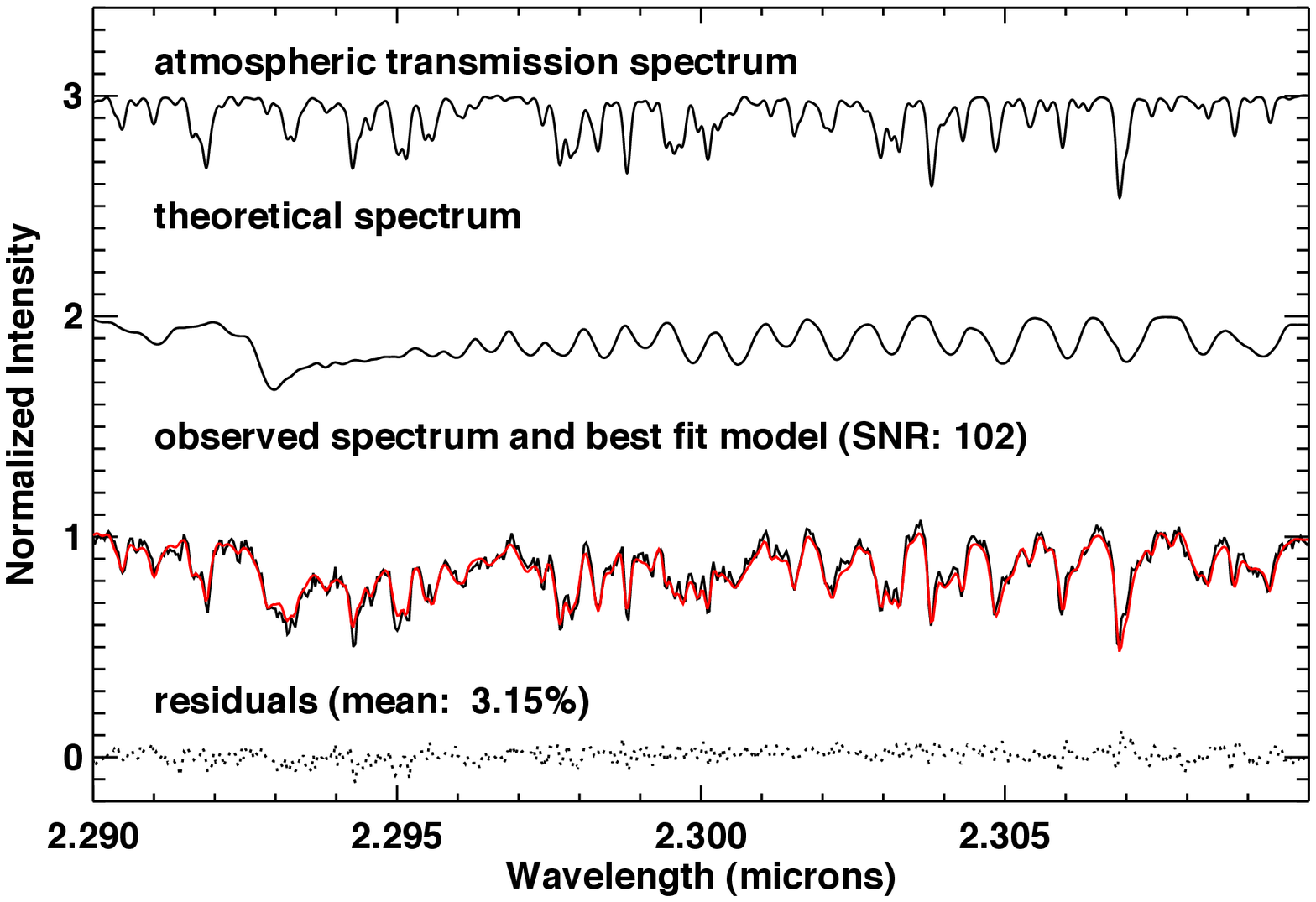}
\figcaption{Plot of the different components involved in modeling the spectrum of the slowly rotating (${vsini}$ = 8 km/s) M3.5V dwarf GJ 752 B (left) and and the rapidly rotating  (${vsini}$ = 61 km/s) M7.5V dwarf 2M1757+70 (right) for the purpose of determining the radial velocity of the star.  They include the telluric spectrum (top) extracted from a high resolution solar spectrum, the spectrum of the target star as
determined from NextGen models (middle), and the combination of these as well as an assumed v{\it sini} and instrument profile (bottom). The bottom dotted line is the
difference between the best fitting model and the data.  \label{finalfit}}
\end{figure}

To obtain the optimal fit to the observed M dwarf spectra, we vary the wavelength solution parameters, the FWHM of the instrumental spectral profile, and the radial velocities and normalization parameters. The fitting is stopped when the $\chi^2$ value
ceases to change between variable perturbations by more than 10$^{-3}$\%. Figure~\ref{finalfit} shows the best fit to one epoch of the spectra for the slowly rotating M3.5 and rapidly rotating M7.5 dwarfs, GJ 752 B and 2M1757+70. The telluric spectrum and the theoretical stellar spectrum rotationally broadened and matched to the published stellar spectral type are also plotted. 

In some cases, a disagreement in the estimated radial velocity determined for the A and B nod pairs ($>$ 100 m/s) during a single epoch requires the manual perturbation of the initial guesses to the fitting parameters in order to bring the two nod RV values into agreement. This additional step was necessary because the amoeba algorithm is sensitive to the initial guesses. We adopted a methodology that automatically finds the appropriate initial guesses but still requires a small amount of customized adjustment by hand to find the best fit to the data. For each epoch of observation we calculate the average of the radial velocities determined from all spectral nod pairs. As described in Bailey et al. (2012), the observed RV precision over multiple epochs is the combination of the theoretical photo noise error, NIRSPEC's instrumental error and the error caused by intrinsic stellar activity.  The theoretical photon noise error derived assuming a perfect instrument, no intrinsic stellar jitter, the signal-to-noise of the observed spectrum and v{\it sini} of the star (Butler et al. 1996). Our NIRSPEC observations of the RV standards (GJ 628, GJ 725 A and GJ 725 B) result in RV precision of 75-143 m/s for nearby, bright, slowly rotating mid-M stars (Bailey et al. 2012). Observations of these same RV standards with greater instrumental precision suggest that they have an intrinsic stellar noise error of $<$ 10 m/s.  The uncertainties quoted for a single RV measurement (see Table~\ref{rvs}) are then the quadrature sum of the photon noise error (varies by star), instrumental error (103 m/s, the average of the RMSs from the RV standards) and stellar jitter error (10 m/s). In Table~\ref{rms} we also include the error associated with each data point ($\sigma_{\chi^2=1}$) that would produce a $\chi^2$ of one if there was no detectible planet around the star resulting in a flat RV curve. 

\section{Results}

The observed standard deviations of the RV measurements for those stars with at least three epochs of observations are listed in Table~\ref{rms} and range from 45 to 2015 m/s for those seven stars in our primary sample of 23 late-type M dwarfs with three or more observations. There are two stars (2M2313+21 and 2M0027+22) that have RV dispersions greater than 1000 m/s and two stars (2M0320+18 and 2M1757+70) with RV 
dispersions greater than 500 m/s. While the fact that one star has an RV dispersion of 45 m/s (2M0253+16) could represent the optimal precision available with this observing method, this comes from only three individual measurements and should be considered tentative. 

The M6 star, 2M2313+21, has the largest scatter in its RV measurements (2015 m/s) and an intermediate rotational velocity (16 km/s). However, with only three RV observations, its difficult to conclude whether the large change in the velocity of the star from 2006 to 2010 is due to an unseen companion. The same issue is also true for 2M0320+18 which has an intermediate observed rotational velocity of 15 m/s and an RV dispersion of 546 m/s and 2M0027+22 which has the highest observed rotational velocity (60 km/s) of all the stars in the sample and an RV dispersion of 1050 m/s. The combination of too few RV measurements, late-type spectral types and large ($>$ 1 km/s) rotational velocities makes the classification of any of these stars as companion candidate premature in the absence of additional RV measurements and a larger sample size. 
Additional RV measurements will allow us to discern between RV variability due to companions, starspots or the measurement floor as a function of stellar spectral type from the NIRSPEC telluric calibrations. 

The M7.5 dwarf, 2M1757+20, has six RV epochs measured over the span of four years (see Table~\ref{rms}). From these observations we see a standard deviation in the RV measurements of 573 m/s (see Figure~\ref{rvs}). Using the theoretical RV precision, number and cadence of the NIRSPEC observations, we utilize Monte Carlo simulations to place limits on the presence of planets around 2M1757+20. Because this survey is most sensitive to short period companions, we set limits at representative orbital periods of 10 and 100 days.  A set of 10,000 circular orbits are generated with random inclinations, phases, and companion masses ranging from $\sim$1 to $80$ Jupiter masses. Circular orbits are a valid assumption for the 10 day orbits if we are assuming that the planets were subject to circularization via disk interactions (Lin et al. 1996). For the 100 day orbits, the addition of orbits over multiple eccentricities effect our mass limits in a way that is smaller that our current measurement uncertainties. In combination with the stellar masses listed in Table \ref{msample}, these orbits are used to produce a set of RV measurements with the same temporal sampling as our NIRSPEC observations. The values of the model RV measurements are perturbed by an error value randomly chosen from a Gaussian distribution with a FWHM of 53 m/s. This observational error corresponds to the median error due to the photon noise and instrument error. Conservative detection limits are then set by determining the companion mass that induces an RV dispersion that is 2$\sigma$ greater than the photon error of the star depending on its rotational velocity. Using a maximum RV amplitude of 573 m/s for 2M1757+20 and assuming a circular orbit, the companion responsible for this perturbation would have a mass of  $<$16 $M_J${\it sini} in a 10 day orbit and $<$40 $M_J${\it sini} in a 100 day orbit. These limits do not rule out that the companion could be a brown dwarf or very-low mass star. 

\subsection{GJ 752 B}

Astrometric measurements of the late-M star GJ 752B (a.k.a. VB 10)  by Pravdo \& Shaklan (2009)  caused great excitement as they showed perturbations that were interpreted as an unseen planet with a mass of 6.4 M$_J$  and period of 0.74 years. This result was touted as the first exoplanet found solely through astrometric measurements. However, the discovery was short-lived as subsequent  radial velocity measurements quickly proved the planet to be non-existent (Bean et al. 2009) and more recent astrometric observations provided further evidence that the original astrometric detection suffered from systematic errors (Lazorenko et al. 2010). Our observations continue the tradition as we observed GJ 752 A and B eight times from June 4 2009 to October 31 2009  (see Table~\ref{rvs}). The dispersions of our nightly RV measurements are 143 m/s and 351 m/s for GJ 752 A and GJ 752 B, respectively (see Figure~\ref{rvs}). The scatter in our NIRSPEC RV measurements are smaller than the measurements of Zapatero Osorio et al. (2009) which reached an amplitude of 1 km/s. Our RV dispersion for GJ 752 B is larger than the 10 m/s reached by Bean et al. (2009) and could be due the smaller spectral resolution of NIRSPEC compared to CRIRES (R=24000 for NIRSPEC versus R=50000 for CRIRES). Monte Carlo simulations of our RV measurements place limits on the mass of a planet around GJ 752 B of 7 and 14 M$_J$ for periods of 10 and 100 days, respectively. 
 
\subsection{Rotational Velocities of Late-M dwarfs}

\begin{figure}[ht]
\epsscale{0.75}
\plotone{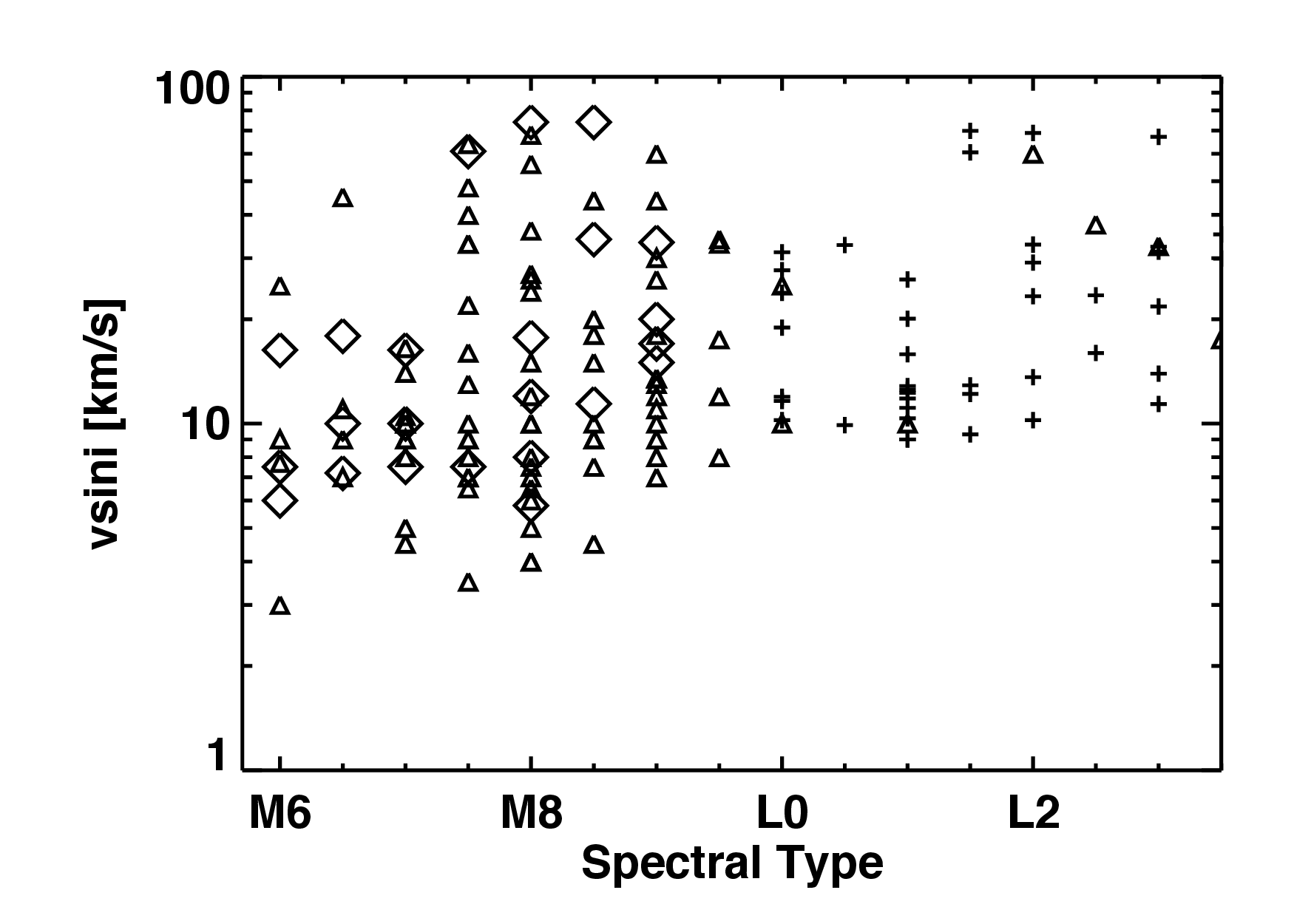}
\figcaption{Plot of our estimated rotational velocities as a function of stellar spectral type (large diamonds) along with those determined for a sample of early L dwarfs also with NIRSPEC (Blake et al. 2010, plusses) and additional late-M dwarfs with v$sini$s determined from optical echelle spectra (Mohanty \& Basri 2003, Reiners \& Basri 2010, triangles). Our data agrees with previous studies in that there appears to be an increase in the rotational velocity with later spectral type. \label{vsinivsspty}}
\end{figure}

Because the stellar rotational velocity is one of the free parameters used when modeling the observed spectrum, we
are able to determine this physical parameter for all of the stars in our late M dwarf sample, which range from 6 to 60 km/s (see Table~\ref{rms}). Figure~\ref{vsinivsspty} plots the values of the rotational velocities estimated from our infrared spectra as a function of spectral type along with those values for a sample of early-L dwarfs also observed with NIRSPEC (Blake et al. 2010) as well as late-M dwarfs observed at optical wavelengths (Mohanty \& Basri 2003; Reiners \& Basri 2010). The uncertainty in the rotation velocity is determined by running the $\chi^2$ minimization program while only allowing the {\it vsini} to vary. Our stated uncertainties are then the standard deviation of all the resulting rotational velocities that provide the best fit to the observed spectra. For those stars with only one epoch of observations (two spectra), the uncertainty stated is the half range of the {\it vsini} values. Our infrared {\it vsini} values agree with those estimated from optical spectra and recently published J (1.25 $\micron$) NIRSPEC spectra (Reiners  \& Basri 2010; Deshpande et al. 2012, see Table~\ref{rms}).  When we combine our {\it vsini} values with those from the literature (see Figure~\ref{vsinivsspty}), our values lie squarely within the spread of values from previous studies (Blake et al. 2010; Mohanty \& Basri 2003; Reiners \& Basri 2010), which show a trend of increasing minimum rotational velocity with later spectral type. 

\section{Discussion and Conclusions}

While we are not able to confirm the detection of any planetary or stellar mass companions with this pilot study of 23 late-M dwarfs with the NIRSPEC instrument on Keck, we have demonstrated that we are capable of
producing RV dispersions of down to 50 m/s for late-M dwarfs with intermediate rotational velocities and telluric wavelength calibrations. Additional stars and RV measurements are needed to definitively conclude whether
those stars with large RV dispersions have unseen companions or whether we are encountering star spots or the noise floor for these spectral types and instrument configuration. 
With radial velocity precisions down to $\sim$50-150 m/s for stars of masses of 0.08-0.15 M$_\odot$, these observations would be sensitive to planets with masses of $<$1 M$_J$sini at periods of 10-100 days. 
Considering there are a few thousand known M dwarfs within 25 parsecs of which a few hundred have V-Ks colors suggesting they are M5 spectral type and later (T. Henry private communication), and that these stars represent a population of stars largely unexplored by most planet detection techniques, these observations open the door to potentially fruitful new discoveries. While the percentage of early-M dwarfs with gas giant planets is currently estimated at 3\%  (Johnson et al. 2010), we have tentative information that this percentage could be $<$ 1\% for late-type M dwarfs at the smallest separations of a$<$0.05 AU (Blake et al. 2010) and little information on these planetary populations at larger separations. With the masses of these stars approaching that of the most massive brown dwarfs, the planet population around them could have significant implications on the method of planet formation that is dominant in this stellar mass regime. 

Saar et al. (1998) and others have used starspot models and observations to show that the radial velocity jitter of a star is directly related its rotational velocity. Therefore, determining the rotational velocities for a large portion of all the of the nearby M dwarfs and determining whether there is any observed correlation between stellar rotational velocity and optical or infrared radial velocity precision would greatly benefit future planet search efforts. 

\begin{figure}[ht]
\epsscale{0.5}
\plotone{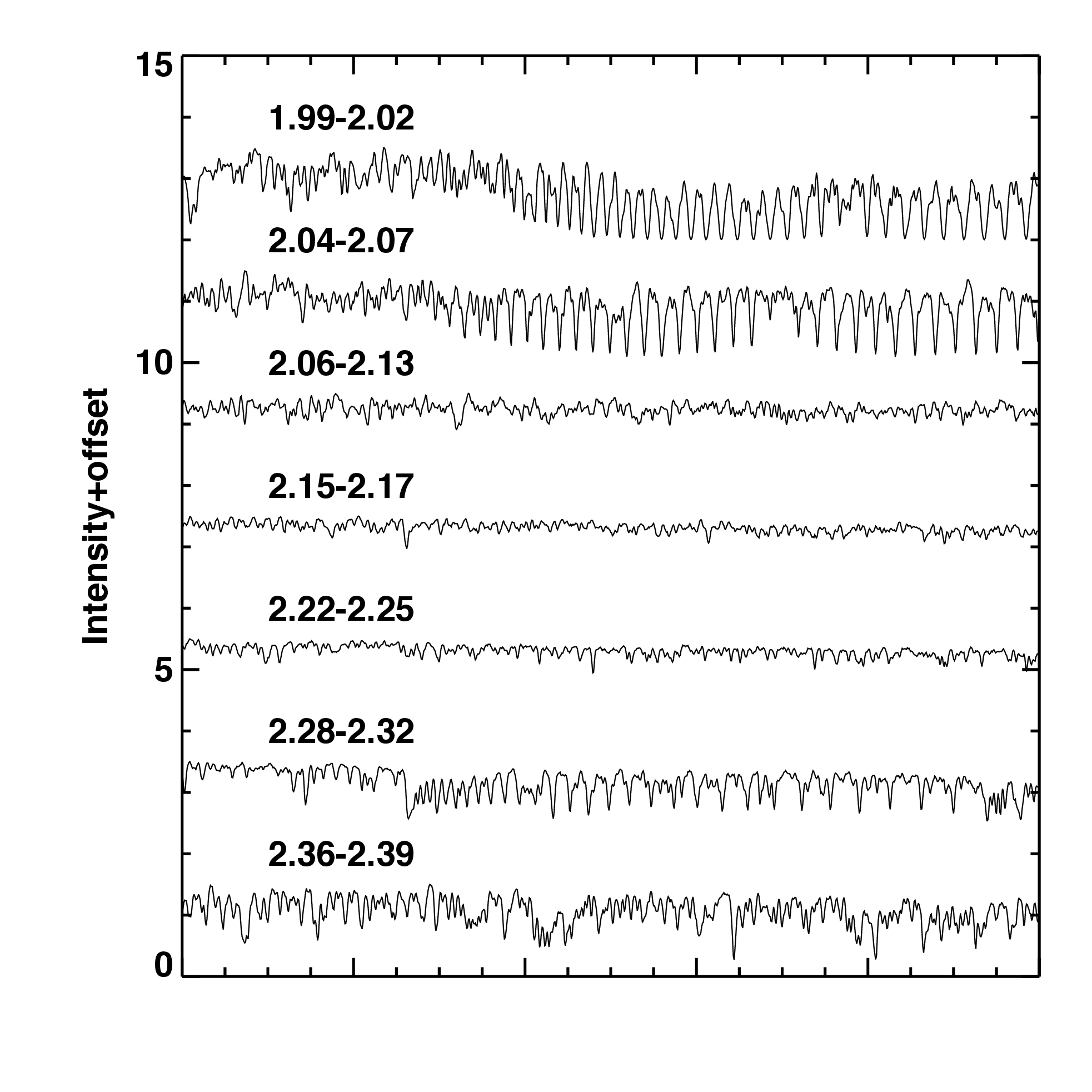}
\figcaption{Plot of all orders of GJ 752 B observed (with wavelengths labeled in microns) with the NIRSPEC-7 filter. In addition to using the  2.28-2.32 $\micron$ region for our
RV determinations we can also use this data to search for Br $\gamma$ emission and conduct spectral
modeling to determine the properties of the atmosphere of these stars. These spectra still contain features due to telluric absorption.   \label{allorders}}
\end{figure}

These high-resolution NIRSPEC spectra can be used for more than just radial velocity and {\it vsini} measurements. Our data represent a additional component of a small sample of K-band, echelle spectra of late-M dwarfs published to date (i.e. Blake et al. 2010). While our radial velocity program uses just the 2.28-2.31 $\micron$ order, the NIRSPEC data cover  a wavelength range of 1.98-2.39 $\micron$ (see Figure~\ref{allorders}). With the discovery of brown dwarfs a few decades ago, modeling the spectral features and, therefore, the complex atmospheres of late-M dwarfs has not been as thorough over all near-infrared bands when compared to what has been accomplished for solar-type stars and brown dwarfs.  None of our spectral show signs of Br $\gamma$ emission (2.166 $\micron$) which could be used to aid in age determinations, however, fully reduced versions of our A star and M dwarf spectra are available to any who wish to utilize them for spectral synthesis and RV templates. 

With so few targets in this pilot study, it is difficult to determine whether there is any observational correlation between radial velocity  precision and the stellar rotational velocity or spectral type of the star. However, we have demonstrated that, with additional observational epochs and a larger sample size, we could detect hot-Jupiter planets around late-M dwarf stars, which are often too faint to be observed efficiently with optical echelle instruments. Planned upgrades to the NIRSPEC instrument ,which include adding a methane gas absorption cell to improve wavelength calibration, could lead to radial velocity precisions of $\sim$20-30 m/s (Bean et al. 2009). Such observations would be sensitive to Jupiter-mass planets in orbits of less than a year around late-M dwarfs. With an emphasis in detecting Earth-like planets set forth by the 2010 Decadal Survey and in searching for planets around M dwarfs set forth by the 2008 Exoplanet Task Force, radial velocity surveys with infrared echelle spectrometers will have a significant impact on future exoplanet studies. 

\begin{acknowledgements}
This research has made use of the Keck Observatory Archive (KOA), which is operated by the W. M. Keck Observatory and the NASA Exoplanet Science Institute (NExScI), under contract with the National Aeronautics and Space Administration. The authors wish to recognize and acknowledge the very significant cultural role and reverence that the summit of Mauna Kea has always had within the indigenous Hawaiian community.  We are most fortunate to have the opportunity to conduct observations from this mountain. We would like to thank the anonymous referee for their insightful comments which improved this manuscript. 
\end{acknowledgements}

\newpage

\begin{deluxetable}{lrrcc}											
\footnotesize											
\tablecaption{Sample of Late M dwarfs \label{msample}}											
\tablehead{ \colhead{Target} & \colhead{RA} & \colhead{Dec} & \colhead{SpTy} & \colhead{K$_s$} }			
\startdata											
2M0027+22	&	00 27 55.9	&	+22 19 32.8	&	M8.0	&	9.57	\nl
2M0140+27	&	01 40 02.6	&	+27 01 50.6	&	M8.5	&	11.43	\nl
2M0227-16	&	02 27 10.4	&	-16 24 48.0	&	M9.0	&	12.14	\nl
2M0253+16	&	02 53 00.9	&	+16 52 53.3	&	M6.5	&	7.59	\nl
2M0253+27	&	02 53 20.3	&	+27 13 33.2	&	M8.0	&	11.48	\nl
2M0306-36	&	03 06 11.6	&	-36 47 52.9	&	M8.5	&	10.63	\nl
2M0320+18	&	03 20 59.7	&	+18 54 23.3	&	M9.0	&	10.64	\nl
2M0752+16	&	07 52 23.9	&	+16 12 15.7	&	M7.0	&	9.85	\nl
2M0818+23	&	08 18 58.1	&	+23 33 52.2	&	M7.0	&	11.15	\nl
2M0853-03 	&	08 53 36.2	&	-03 29 32.1	&	M9.0	&	9.94	\nl
2M1049+25	&	10 49 41.5	&	+25 38 53.6	&	M6.0	&	11.44	\nl
2M1124+13	&	11 24 53.3	&	+13 22 53.4	&	M6.5	&	10.07	\nl
2M1546+37	&	15 46 05.4	&	+37 49 45.8	&	M7.0	&	11.41	\nl
2M1707+64	&	17 07 18.3	&	+64 39 33.1	&	M9.0	&	11.38	\nl
2M1757+70	&	17 57 15.4	&	+70 42 01.2	&	M7.5	&	10.40	\nl
2M1835+32	&	18 35 37.9	&	+32 59 54.6	&	M8.5	&	9.17	\nl
GJ752 B$^a$	&	19 16 57.6	&	+05 09 02.0  	&	M8.0	&	8.77	\nl
2M2052-23	&	20 52 08.6	&	-23 18 09.6	&	M6.5	&	11.29	\nl
2M2221+27	&	22 21 54.4	&	+27 29 07.0 	&	M6.0	&	11.52	\nl
2M2235+18	&	22 35 49.1	&	+18 40 29.9	&	M7.0	&	11.37	\nl
2M2306-05	&	23 06 29.3	&	-05 02 28.6	&	M7.5	&	10.30	\nl
2M2313+21	&	23 13 47.4	&	+21 17 29.4 	&	M6.0	&	10.44	\nl
2M2349+12	&	23 49 49.0	&       +12 24 39.0       &	M8.0	&	11.56	\nl
\enddata		
\tablenotetext{a}{aka VB 10}									
\end{deluxetable}

\begin{deluxetable}{lllcr}											
\footnotesize											
\tablecaption{Calibration Stars \label{sample}}											
\tablehead{ \colhead{Target} & \colhead{RA (2000)} & \colhead{Dec (2000)} & \colhead{SpTy} & \colhead{K$_s$} 
}	
\startdata	
A stars  & & & & \nl						
\hline			
HR 104   & 00 28 13.7  &  +44 23 40.0 & A2 & 9.99 \nl
HR 8518 & 22 21 39.4  & -01 23 14.4 & A0 & 4.02  \nl 
HR 5511 & 14 46 14.9  &  +01 53 34.4 & A0 & 3.65 \nl
HR 7724 & 20 14 16.6  & +15 11 51.4  & A0 & 4.77 \nl 
HR 7336 & 19 20 35.7  & -00 53 31.8 & B9 IV/V & 5.46 \nl
RV Standards & & & & \nl
\hline
GJ 628     & 16 30 18.1 & -12 39 45.3 & M3.5 & 5.08  \nl
GJ 725 A & 18 42 46.7 & +59 37 49.4   & M3.0    & 4.43 \nl
GJ 725 B & 18 42 46.9 & +59 37 36.7   & M3.5   &  5.00 \nl
GJ 752 A & 19 16 55.3 & +05 10 08.1   & M3.5 & 4.67 \nl 
\enddata											
\end{deluxetable}	
	
\begin{center}	
\begin{deluxetable}{lccc}							
\footnotesize							
\tablecaption{Radial Velocity Measurments \label{rvs}}							
\tablehead{ \colhead{Star} & \colhead{HJD$^a$ - 2400000} & \colhead{Date [Y.M.D]}  & \colhead{RV [m/s]}}							
\startdata																	
2M0027+22	&	53928.576	&	2006.07.12	&	-9633$\pm$181	\nl
	&	55401.559	&	2010.07.24	&	-11428$\pm$226	\nl
	&	55402.555	&	2010.07.25	&	-11475$\pm$202	\nl
\hline	&		&		&		\nl
2M0253+16	&	53928.620	&	2006.07.12	&	68330$\pm$107	\nl
	&	55401.630	&	2010.07.24 	&	68253$\pm$106	\nl
	&	55402.637	&	2010.07.25 	&	68250$\pm$102	\nl
\hline	&		&		&		\nl
2M0320+18	&	53931.633	&	2006.07.15 	&	45953$\pm$118	\nl
	&	55401.610	&	2010.07.24	&	45638$\pm$111	\nl
	&	55402.621	&	2010.07.25 	&	44890$\pm$104	\nl
\hline	&		&		&		\nl
2M1707+57	&	53929.415	&	2006.07.13	&	-10001$\pm$151	\nl
	&	54308.372	&	2007.07.27	&	-9990$\pm$185	\nl
	&	54309.390	&	2007.07.28	&	-11184$\pm$149	\nl
	&	54311.367	&	2007.07.30	&	-11042$\pm$150	\nl
	&	54312.428	&	2007.07.31	&	-11153$\pm$147	\nl
	&	55401.452	&	2010.07.24	&	-11010$\pm$181	\nl
\hline	&		&		&		\nl
2M2306-05	&	53929.583	&	2006.07.13	&	-52919$\pm$108	\nl
	&	55401.536	&	2010.07.24 	&	-52600$\pm$106	\nl
	&	55402.581	&	2010.07.25	&	-52746$\pm$106	\nl
\hline	&		&		&		\nl
2M2313+21	&	53929.562	&	2006.07.13	&	-7391$\pm$116	\nl
	&	55401.582	&	2010.07.24	&	-10873$\pm$117	\nl
	&	55402.602	&	2010.07.25	&	-10889$\pm$118	\nl
\hline	&		&		&		\nl
GJ 752b	&	55016.448	&	2009.07.04	&	35614$\pm$104	\nl
	&	55024.496	&	2009.07.12	&	35644$\pm$103	\nl
	&	55041.259	&	2009.07.29	&	35738$\pm$102	\nl
	&	55110.281	&	2009.10.06	&	36483$\pm$102	\nl
	&	55115.297	&	2009.10.11	&	36161$\pm$102	\nl
	&	55116.315	&	2009.10.12	&	36169$\pm$104	\nl
	&	55135.202	&	2009.10.31	&	36491$\pm$115	\nl
	&	55401.510	&	2010.07.24	&	36081$\pm$105	\nl
\enddata
\tablenotetext{a}{HJD - heliocentric Julian date. All measurements heliocentric radial velocities.}							
\end{deluxetable}	
\end{center}
			
\begin{center}
\begin{deluxetable}{lcccccc}										
\footnotesize										
\tablecaption{RV RMS and v{\it sini} Values for RV Standards and Science Targets  \label{rms}}											
\tablehead{ \colhead{Target} & \colhead{N}   & \colhead{RV}  & \colhead{RV RMS} & \colhead{ $\sigma_{\chi^2=1}$ } & \colhead{v{\it sini} - this work}  &  \colhead{v{\it sini} - published$^a$} \nl													
                                                    &                          & \colhead{km/s} & \colhead{m/s} &  \colhead{m/s}  &\colhead{km/s}  & \colhead{km/s}													
}													
\startdata													
GJ 628          	&	10	&	-21.1	&	80	&	 ... 	&	 ... 	&	1	\nl
GJ 725 A      	&	5	&	-0.5	&	89	&	 ... 	&	 ... 	&	1	\nl
GJ 725 B      	&	13	&	1.4	&	75	&	 ... 	&	 ... 	&	1	\nl
GJ 752 A      	&	5	&	36.2	&	143	&	 ... 	&	 ... 	&	2	\nl
\hline													
GJ 752 B      	&	8	&	36.0	&	351	&	350	&	  8$\pm$7	&	8	\nl
2M0027+22	&	3	&	-10.8	&	1050	&	820	&	60$\pm$12	&	56	\nl
2M0140+27	&	1	&	10.4	&	 ... 	&	...	&	7$\pm$4	&	$<$12	\nl
2M0227-16	&	2	&	49.0	&	 ... 	&	...	&	17$\pm$11	&	 ... 	\nl
2M0253+16	&	3	&	68.3	&	45	&	37	&	10$\pm$4	&	 ... 	\nl
2M0253+27	&	1	&	47.8	&	 ... 	&	...	&	18$\pm$5	&	 17.5 	\nl
2M0306-36	&	2	&	12.2	&	 ... 	&	...	&	10$\pm$7	&	 ... 	\nl
2M0320+18	&	3	&	45.5	&	546	&	430	&	15$\pm$6	&	15	\nl
2M0752+16	&	1	&	-15.5	&	 ... 	&	...	&	6$\pm$4	&	 ... 	\nl
2M0818+23	&	1	&	33.5	&	 ... 	&	...	&	8$\pm$3	&	 ... 	\nl
2M0853-03 	&	1	&	10.3	&	 ... 	&	...	&	20$\pm$7	&	 ... 	\nl
2M1049+25	&	1	&	7.3	&	 ... 	&	...	&	8$\pm$5	&	 ... 	\nl
2M1124+13	&	1	&	8.4	&	 ... 	&	...	&	7$\pm$4	&	 ...	\nl
2M1546+37	&	1	&	-22.5	&	 ... 	&	...	&	10$\pm$7	&	10	\nl
2M1707+64	&	2	&	-10.1	&	 ... 	&	...	&	33$\pm$12	&	 ...	\nl
2M1757+70 	&	6	&	-12.3	&	573	&	500	&	34$\pm$11	&	 33	\nl
2M1835+32	&	2	&	4.8	&	 ... 	&	...	&	34$\pm$17	&	44	\nl
2M2052-23	&	1	&	35.8	&	 ... 	&	...	&	18$\pm$10	&	 ...	\nl
2M2221+27	&	1	&	-14.8	&	 ... 	&	...	&	6$\pm$4	&	 ...	\nl
2M2235+18	&	1	&	-14.9	&	 ... 	&	...	&	16$\pm$8	&	 ...	\nl
2M2306-05	&	3	&	-52.8	&	160	&	130	&	  8$\pm$5	&	6	\nl
2M2313+21	&	3	&	-9.3	&	2015	&	1600	&	16$\pm$4	&	 ...	\nl
2M2349+12	&	1	&	-2.0	&	 ... 	&	...	&	12$\pm$4	&	 $<$12	\nl
\enddata																					
\tablenotetext{a}{Reiners \& Basri 2009 (optical) and Deshpande et al. 2012 (infrared).}								
\end{deluxetable}		
\end{center}							
			
%%%%%%%%%%%%%%%%%%%%%%%%%%%%%
%%%%%%%%%%%%%%%%%%%%%%%%%%%
%%%%%%%%%%%%%%%%%%%%%%%%%%%%%
%%%%%%%%%%%%%%%%%%%%%%%%%%%
%%%%%%%%%%%%%%%%%%%%%%%%%%%%%
\begin{figure}[ht]
\epsscale{0.75}
\plotone{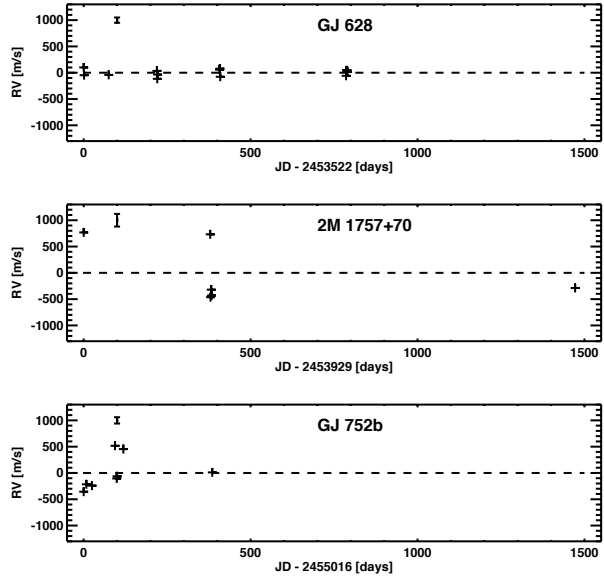}
\figcaption{Top: Plot of the RV measurements for the standard star GJ 628. Middle and Bottom: Plots of the RV measurements for GJ 752 B and 2M1757+70. The error bar on the top, left corner of each plot is scaled to the average of the errors of each of the RV measurements which are derived from the photon errors (50 - 100 m/s). \label{rvs}}
\end{figure}
%%%%%%%%%%%%%%%%%%%%%%%%%%%
%%%%%%%%%%%%%%%%%%%%%%%%%%%%%
%%%%%%%%%%%%%%%%%%%%%%%%%%%

\end{document}